# A dynamic mathematical model for sub-lethal damage recovery: Integrating recovery processes into multi-target theory to visualize damage states

Renta Shibata[a], Fuminobu Sato[a], Yoko Akiyama[a], Yuichiro Manabe[a*]

[a]*Division of Sustainable Energy and Environmental Engineering, Graduate School of Engineering, The University of Osaka, 2-1 Yamadaoka, Suita, Osaka 565-0871, Japan*

Corresponding author E-mail address: [manabe_y@see.eng.osaka-u.ac.jp]

**ABSTRACT**

Quantifying the biological effects of radiation is essential for radiotherapy and radiation protection. While research has centered on statistical analyses of experimental data, this approach often fails to elucidate underlying mechanisms, limiting the development of generalizable models. Describing time-dependent phenomena, such as sub-lethal damage (SLD) recovery during fractionated irradiation, presents a particularly significant challenge for existing static models. To address this, we developed a mechanistic mathematical model that extends the classical multi-target, single-hit model by introducing explicit recovery transitions among damage states, allowing the simultaneous description of radiation-induced damage accumulation and biological recovery. We applied this model to Elkind and Sutton's classical split-dose experimental data to determine its parameters. The model reproduced the major features of split-dose survival curves for treatments with 0-, 2.5-, and 23-hour recovery intervals. We visualized the temporal changes in damage state distributions within the cell populations. These simulations provided a process-level interpretation of the characteristic shoulder-like response appearing after an inter-fraction recovery interval. Furthermore, the estimated recovery half-life was ~120 minutes, which is broadly consistent with reported timescales of DNA double-strand break repair in mammalian cells, supporting the biological interpretability of the model parameters. The proposed framework provides a simple, interpretable description of radiation-induced cell death dynamics, including SLD recovery, and may serve as a basis for the improved mechanistic modeling of radiobiological effects under different irradiation protocols.

## 1. Introduction

A quantitative understanding of the biological effects of radiation necessitates mechanistic mathematical models that describe the interaction between physical damage induction and biological repair using time-dependent ordinary differential equations. The dose-rate effect, arising from the competition between damage and recovery, is a particularly crucial aspect of this relationship, as it forms the basis for radiation protection and treatment planning, requiring an explainable, mechanistic framework. International assessments have emphasized the importance of incorporating recovery processes when estimating effects at low doses and low dose rates (Rühm et al., 2016). This study aims to develop such a model.

Historically, Elkind and Sutton (1959, 1960) demonstrated recovery from sub-lethal damage (SLD) in fractionated irradiation experiments, quantifying a phenomenon in which a "shoulder-like response" appears following subsequent irradiation. Their seminal papers in *Nature* (1959) and *Radiation Research* (1960) established the classical experimental finding that recovery proceeds rapidly and that a sufficient interval between doses allows significant survival. This landmark dataset still serves as a benchmark for comparing models that test time-dependent recovery processes. Therefore, this study focuses on the Elkind data for validation.

While conventional target theory concisely explains the shape of survival curves, it does not explicitly include a recovery process, making it difficult to link recovery parameters to biological mechanisms. Lea and Catcheside (1942) introduced a correction factor to account for recovery effects during irradiation, but this did not directly describe the recovery that occurs during the period between doses in fractionated irradiation. Conversely, the linear-quadratic (LQ) model is widely used in clinical applications, but its parameters have limited mechanistic interpretability (Yoshida et al., 2019). Reviews in the clinical and protection fields, while positioning the LQ model as the standard, conclude that it must be supplemented with mechanistic models, particularly for handling dose-rate and time effects. Thus, a theory that integrates recovery into the system of equations is needed (Fowler, 1989).

Representative examples of mechanistic models include the Lethal–Potentially Lethal (LPL) model (Curtis, 1986), the Repair–Misrepair–Fixation (RMF) model (McMahon et al., 2019), and the Microdosimetric Kinetic model (MKM) (Hawkins, 1998), which treat damage induction, repair, and misrepair using kinetic descriptors. These models help link DNA double-strand break (DSB) kinetics and radiation quality to cell survival, but a simple and visualizable framework for Elkind-type sub-lethal damage recovery remains useful. Non-homologous end joining (NHEJ) and homologous recombination (HR), with different kinetics and cell-cycle dependencies, are the two major DSB-repair pathways. We therefore develop a compact, ordinary-differential-equation model that adds explicit recovery transitions to target theory.

We previously proposed the Whack-a-Mole theory, which treats dose-rate dependence in principle, and validated it against genetic mutation data (Manabe et al., 2015). The core supposition of this theory is that the observed biological effects are determined by the dynamic balance between two processes: physical damage induction by radiation and biological repair by the organism. In this study, we apply a recovery model that extends this framework to radiation-induced cell death. The basic concept of this model was proposed by Wada et al. (in press), and this study performs its first rigorous validation using Elkind's data. Specifically, we identify the target number, intrinsic sensitivity, and recovery rate constant and evaluate whether the reappearance of a shoulder-like response after inter-fraction recovery intervals of 0, 2.5, and 23 hours, as reported by Elkind and Sutton (1960), can be reproduced using a common parameter set. Furthermore, by visualizing the time evolution of the damage-state distribution, we link population-level survival changes to the underlying effective damage-state dynamics. This provides a pre-clinical and phenomenological basis for investigating time-dependent recovery processes under different irradiation protocols, but direct application to clinical treatment planning will require further validation with appropriate experimental and clinical datasets.

The rest of this paper is structured into five sections: Section 2 describes Elkind and Sutton's classical split-dose experimental dataset; Section 3 presents the mathematical model, which incorporates explicit recovery transitions among damage states into the conventional multi-target, single-hit model; Section 4 applies the model to the dataset, determining model parameters, and visualizes the temporal progression of damage states within the cell population; Section 5 discusses the biological interpretation, theoretical significance, limitations, and future prospects of the model; and, finally, Section 6 summarizes the study's conclusions.

## 2. Dataset

In this study, we use classical split-dose X-ray data for Chinese hamster V79 cells reported by Elkind and Sutton (1960). A first dose of 5.05 Gy was delivered at 7.22 Gy/min, followed by either no interval (0 h) or an incubation interval of 2.5 or 23 h before the second irradiation. Since numerical values were not tabulated in the original article, they were derived from the points in Fig. 11 of Elkind and Sutton (1960), digitized using the Graph Digitizing System (GSYS) (Hokkaido University Nuclear Reaction Data Centre, 2022). The resulting dataset, used for fitting, is shown in the present study's Fig. 1 and is available in the Mendeley Data repository (DOI: 10.17632/3wrx25by4s.1). The intervals represent incubation at 37°C between irradiations, not colony-scoring times. The exact final scoring times for the original experiment were not reported separately; the article states that controls were usually scored after 12 days and irradiated plates after longer incubation periods extending up to 17 days at low survival. Accordingly, our simulation endpoint is the end of the irradiation–recovery–irradiation protocol, i.e., immediately after the second irradiation; subsequent colony growth is treated as an assay stage and is not modeled.

## 3. Mathematical model

In this section, we construct a model that introduces explicit recovery transitions among damage states into the conventional multi-target, single-hit model in which a cell has several "targets" and cell death occurs when all of them receive one or more injuries ("hits").

### *3.1. System of ordinary differential equations for the dynamic model*

At its core, the dynamic process described in this model is a Markov chain process, with the subsequent system of ordinary differential equations describing a Kolmogorov forward equation.

Based on this mathematical foundation, target theory provides a fundamental framework for understanding radiation-induced cell death. The concept underlying target theory represents one of the earliest theoretical attempts to explain radiation's action in tissues, dating back to the pioneering work on the "Point Heat Theory" by Dessauer and its mathematical formalization by Crowther (Dessauer, 1922; Crowther, 1924). Initially, it was a simple model in which an event occurred with a single hit to a single target, but this concept later evolved into multi-target or multi-hit models to explain the "shoulder" shape seen in mammalian cell survival curves (Lea, 1946). In its general form, the multi-target, single-hit model assumes that there are $n$ parts (called "targets") in a cell at which a sub-lethal damage event occurs if hit, and inactivation (here, cell death) occurs if all of them receive one or more hits.

Let $p^{(i)}(t)$ denote the fraction of cells with $i$ damaged targets at time $t$, where $i = 0, 1, \cdots, n$. States $i = 0, 1, \cdots, n-1$ correspond to clonogenically surviving states, whereas $i = n$ is treated as the lethal absorption state. The surviving fraction is therefore defined as $S(t) = \Sigma_{i=0}^{n-1} p^{(i)}(t)$, and $p^{(n)}(t) = 1 - S(t)$ represents the lethally damaged fraction.

Assuming hits occur randomly according to a Poisson distribution, when irradiated at a constant dose rate $\gamma$ [Gy/min], the probability $P(t)$ that a single target is hit by time $t$ is given by $P(t) = 1 - e^{-\kappa\gamma t}$. Here, $\kappa$ $[1/Gy]$ is an intrinsic radiosensitivity parameter representing the hit probability per unit dose.

To describe irradiation and recovery in a unified manner, we write the dose-rate input and recovery rate as $\gamma(t)$ and $\rho(t)$, respectively. In the present analysis, $\gamma(t)$ is a protocol-defined piecewise function: $\gamma(t)$ is the prescribed dose rate during irradiation and 0 during recovery intervals. In our main model, $\rho(t)$ is taken as a constant, $\rho$, during the recovery process. The explicitly time-dependent $\rho(t)$ formulation is considered separately as an exploratory extension.

In the multi-target model, the proportion of cells $p^{(i+1)}(t)$ where $i + 1$ out of $n$ targets have been hit by time $t$ increases in proportion to $p^{(i)}(t)$ and the number of unhit targets in the associated cells $(n - i)$ and decreases in proportion to $p^{(i+1)}(t)$ and the number of unhit targets in these cells $(n - (i + 1))$. Considering transitions from the $i$ and $i + 1$ hit states, the rate of change of the proportion of cells with $i + 1$ hits, $p^{(i+1)}$, is described by a system of ordinary differential equations (the complete transition scheme and notations are illustrated in Fig. 2):

for $i = 0$,

$$\frac{dp^{(0)}(t)}{dt} = -n\kappa\gamma(t)p^{(0)}(t) + \rho(t)p^{(1)}(t);$$

for $i = 1, \cdots, n-2$,

$$\frac{dp^{(i)}(t)}{dt} = \{n - (i-1)\}\kappa\gamma(t)p^{(i-1)}(t) - \{(n-i)\kappa\gamma(t) + i\rho(t)\}p^{(i)} + (i+1)\rho(t)p^{(i+1)}(t);$$

for $i = n-1$,

$$\frac{dp^{(n-1)}(t)}{dt} = 2\kappa\gamma(t)p^{(n-2)}(t) - \{\kappa\gamma(t) + (n-1)\rho(t)\}p^{(n-1)};$$

and for $i = n$,

$$\frac{dp^{(n)}(t)}{dt} = \kappa\gamma(t)p^{(n-1)}(t) \tag{1}$$

In this formulation, the lethal state $p^{(n)}(t)$ is treated as an absorbing state. Therefore, recovery transitions are allowed only among the surviving states, $i = 1, \cdots, n-1$, with no recovery transition from $p^{(n)}(t)$ to $p^{(n-1)}(t)$ included.

For the no-recovery case, the solution to this system of equations under the initial conditions $p^{(0)}(0) = 1$ and $p^{(i)}(0) = 0$ for $i > 0$ is given by

$$p^{(i)}(t) = \binom{n}{i}(1 - e^{-\kappa\gamma t})^i e^{-(n-i)\kappa\gamma t} \tag{2}$$

Cell survival $S(t)$, the proportion of cells with fewer than $n$ targets hit, i.e., $S(t) = 1 - p^{(n)}(t)$, simplifies to

$$S(t) = 1 - (1 - e^{-\kappa\gamma t})^n \tag{3}$$

Equations (2) and (3) are the closed-form solutions for the no-recovery case, $\rho(t) = 0$. For the general case incorporating both damage induction and recovery, Eq. (1) is integrated numerically using the protocol-defined $\gamma(t)$ and $\rho(t)$. During an inter-fraction recovery interval, $\gamma(t) = 0$ and $\rho(t) = \rho$, whereas during the acute-irradiation approximation used for fitting, $\rho(t) = 0$. Thus, Eq. (1) also provides the general recovery-phase solution using the corresponding matrix exponential for constant coefficients or numerical integration.

This system of ordinary differential equations enables the simultaneous progression of damage induction and repair to be described using the parameters $n, \kappa,$ and $\rho$. Unless otherwise stated, survival fractions in simulations were evaluated at the operational simulation endpoint, the end of the irradiation–recovery–irradiation protocol. Thus, for split-dose conditions, evaluation occurred immediately after completion of the second irradiation.

### *3.2. Basic biological assumptions of the model*

This model formulates physical damage induction and biological repair as a system of ordinary differential equations to describe the dynamic process of radiation-induced cell death. To ensure this mathematical framework's biological validity and interpretability, we explicitly set five basic assumptions that are essential for defining the scope and limitations of the model:

1. Effective Damage Events and Their Relation to DSBs
   In this phenomenological model, a hit is interpreted as an effective radiation-induced damage event that contributes to a loss of clonogenic survival. While DSBs are biologically important lesions and likely contribute to such events, we do not assume a one-to-one correspondence between one model hit and one physical DSB. Thus, in this coarse-grained target-theory description, the target number $n$ represents the number of effective damage states or targets required for inactivation, rather than a literal count of DSBs.

2. Homogeneity of Targets
   We assume that all potential targets, interpreted here as effective damage sites, have the same radiosensitivity. This means that the probability of generating an effective damage event per unit dose is treated as being equal for all targets, regardless of genomic location (e.g., euchromatin vs. heterochromatin) or transcriptional status, and that the contribution to cell death is equivalent, regardless of which target is damaged. This approximation keeps the model, which aims to describe the average response of the cell population as a whole, simple.

3. Linear Recovery Kinetics
   Recovery is represented as a first-order transition among effective damage states. A cell in state $i$ returns to state $i - 1$ at a rate of $i\rho(t)$, so each accumulated damage event is assumed to recover independently at the same instantaneous rate. In the main fitting, $\rho(t) = \rho$ is constant during the recovery interval. This coarse-grained assumption is not meant to imply that all molecular repair pathways are identical, but rather that $\rho$ is an effective, population-averaged recovery parameter.

4. Approximate Handling of Acute Irradiation Data (0-hour recovery)
   The dynamic model constructed in this study (Eq. (1)) describes damage induction and recovery simultaneously. The single-dose data (0-hour recovery), used to determine the target number and intrinsic radiosensitivity, correspond to the high dose-rate (7.22 Gy/min) condition in Elkind and Sutton's experiment. The irradiation time for the first 5.05 Gy exposure is approximately 0.70 min, which is short compared with the recovery timescales of the split-dose data. Therefore, the conventional multi-target survival function, which does not include explicit recovery during irradiation, was used as a practical approximation for fitting the acute single-dose data.
   Note that we do not use this assumption to prove that repair during irradiation is absent or biologically negligible. The comparison with simultaneous recovery during the short irradiation period, using a recovery-rate parameter estimated from split-dose recovery data, is treated only as a local sensitivity analysis for the specific high-dose-rate protocol used by Elkind and Sutton. This point is clarified in the Results section.

5. Absence of Cell Proliferation

   We assume that cell proliferation (repopulation) does not occur during the recovery period. This approximation is reasonable for the 2.5-hour interval, given the reported 10–14-hour V79 cell cycle, but may fail for the 23-hour interval. In every protocol, the modeled endpoint is immediately after completion of the second irradiation. The possible contribution of cell proliferation in the 23-hour recovery model is discussed in Section 5.4.2.

*3.3.Model extension: Introduction of a time-dependent recovery rate* $\rho(t)$

As an exploratory phenomenological extension, we allowed the recovery rate to vary with the inter-fraction interval. In this extension, the recovery rate was modeled based on a transient reduction from a baseline value:

$$\rho(t) = \max\left[0, \rho_{base} - A_\rho\, exp\left(-\frac{\left(t - t_{c,\rho}\right)^2}{2{\sigma_\rho}^2}\right)\right], (4)$$

where $\rho_{base}$ is the baseline recovery rate, $A_\rho$ is the amplitude of the transient reduction, $t_{c,\rho}$ is the center time of the transient component, and $\sigma_\rho$ is its characteristic width. An alternative $\kappa(t)$ extension, shown in Fig. 7, was parameterized analogously using $A_\kappa$, $t_{c,\kappa}$, $\sigma_\kappa$, and $\kappa_{base}$, with $\rho_{const}$ held constant. The fitted parameters are reported in Section 4.6, and the experimental data and both exploratory extensions are shown in Fig. 7. These analyses were performed using recovery-time survival data from Fig. 16 of Elkind and Sutton (1960), separately from the main split-dose survival-curve fitting shown in Fig. 3.

## 4. Results

*4.1. Single-dose condition*

The model parameters *n* (number of targets) and *κ* (intrinsic radiosensitivity) were determined using single-dose data (0-hour recovery) from Elkind and Sutton. As stated in Assumption 4 in Section 3.2, the conventional multi-target model based on data from treatments without a recovery term was used as a practical approximation for this high-dose-rate acute irradiation condition. The effect of including recovery during the short irradiation period is examined only as a local sensitivity analysis in Section 4.2.2.

To ensure appropriate fitting across the entire dose range, including the low survival region, the logarithm of the surviving fraction was taken, and the least-squares method was applied to the logarithm of the model-predicted values.

The number of targets $n$ was calculated to be 4.9, so the integer value $n = 5$ was adopted. Fitting was performed again with $n = 5$ fixed, yielding $\kappa = 0.7$ 1/Gy. Since the number of targets is biologically interpretable as the number of sensitive sites within the cell, it is natural to treat it as an integer. When changing $n$ from 4.9 to 5, the residual sum of squares (RSS), denoted here as $\chi^2$, changed only marginally, from 0.2370 to 0.2371, indicating no significant impact on the quality of the fit.

*4.2. Quantitative reproduction of fractionated irradiation survival curves and model goodness-of-fit*

*4.2.1. Reproduction of Elkind curves*

During the recovery interval, no radiation treatment is delivered, and therefore, $\gamma(t) = 0$. Starting from the damage-state distribution $p^{(i)}$ at the end of the first irradiation, recovery dynamics were obtained across the prescribed recovery interval as a special case of Eq. (1) in which $\gamma(t) = 0$ and $\rho(t) = \rho$. The resulting distribution was then used as the initial condition for the second irradiation.

Elkind and Sutton's experimental data include two recovery periods: a 2.5-hour recovery and a 23-hour recovery. Although the periods differ, the essential speed of recovery was treated identically. Therefore, the recovery rate constant $\rho$ was determined as a common parameter for both fractionated irradiation datasets by fitting the split-dose predictions obtained from Eq. (1) using the irradiation–recovery–irradiation protocol. The obtained value of $\rho$ was $0.006\,[1/\mathrm{min}]$. The Python "lmfit" library, a non-linear least-squares library, was used for fitting. Fig. 3 shows the reproduction of Elkind and Sutton's experimental data (points) and the model curves based on the determined parameters for each condition. The estimated model parameters and fixed protocol inputs used in the simulations are summarized in Table 1.

We evaluated the model residuals using all 22 log-transformed observations across the 0-hour, 2.5-hour, and 23-hour conditions. For split-dose treatments, the results are shown only from the first 5.05 Gy dose onward; no extrapolation below a total dose of 5.05 Gy is shown in Fig. 3.

The adjusted coefficient of determination ($R^2_{\mathrm{adj}}$), corrected for the number of model parameters (three: $n$, $\kappa$, and $\rho$), was 0.9850, a value close to 1, indicating that the model explains approximately 98.5% of the variance. The residual sum of squares (RSS) was 1.3287. As RSS is dimensionless on the survival data's logarithmic scale, its magnitude is best interpreted relative to the total variance of the dataset. The corresponding root mean square error (RMSE) was approximately 0.264. These diagnostics indicate that the model captures the dominant trend of the Elkind–Sutton dataset, including the reappearance of a shoulder-like response after the inter-fraction recovery interval. Nevertheless, systematic deviations can remain in the low-dose region because the underlying multi-target structure has a zero initial slope for an initially undamaged population; this structural limitation is discussed below.

Descriptive diagnostics were $N = 9$, RSS = 0.300, and RMSE = 0.182 for 0-hour recovery data; $N = 6$, RSS = 0.034, and RMSE = 0.075 for 2.5-hour recovery data; and $N = 7$, RSS = 0.995, and RMSE = 0.377 for 23-hour recovery data (all residuals on the natural-log scale). The individual RSS values sum to the reported overall RSS of 1.329. No residual cutoff or condition-specific hypothesis test was prespecified, so the summaries are descriptive rather than pass/fail criteria. The larger discrepancy seen in the 23-hour recovery data is consistent with limitations of the no-proliferation assumption, whereas the zero-initial-slope property remains a structural limitation of the model for an initially undamaged population at low doses.

### *4.2.2. The effects of allowing recovery during the short irradiation period*

We next performed a local sensitivity analysis to examine the effect of including a recovery term during the short, high-dose-rate irradiation period used in the Elkind–Sutton experiment. This analysis was not intended to prove that biological repair during irradiation is generally negligible. Rather, it was intended to assess the magnitude of the difference between simulations with and without this added variable. Fig. 4 compares the survival curve calculated without recovery during irradiation (Eq. (3)) with the curve calculated using the full system including simultaneous recovery during irradiation (Eq. (1)).

As $\rho$ was estimated from the split-dose data, this comparison is a sensitivity check rather than independent evidence that repair during irradiation is absent.

At 5.05 Gy, including simultaneous recovery increased predicted survival by 0.95% at 7.22 Gy/min, while at 1, 2, 3, 4, and 5 Gy/min, survival increased by 6.89%, 3.45%, 2.30%, 1.72%, and 1.38%, respectively. Thus, the effect remained small near the high experimental dose rate but increases as the dose rate decreases. We, therefore, restricted the conclusion to Elkind and Sutton's results and cannot generalize the negligible effects of including simultaneous recovery to low-dose-rate irradiation.

### *4.3. Temporal evolution of intracellular damage states*

This model can predict not only the final surviving fraction but also the temporal evolution of the damage-state distribution $p^{(i)}$ within the cell population. Fig. 5 shows the temporal evolution of damage states for the 0-hour, 2.5-hour, and 23-hour recovery conditions.

Under split-dose protocols, during the first irradiation phase, $p^{(0)}(t)$ rapidly decreased while the lethally damaged state $p^{(n)}(t)$ increased substantially. This is due to the high dose rate of the exposure. In the 0-hour recovery treatment, the second irradiation begins immediately, so the damage-state distribution remained strongly shifted toward highly damaged states at the start of the second irradiation.

In contrast, during the 2.5-hour and 23-hour recovery intervals, the distribution of surviving cells shifted from higher damage states toward lower damage states as repair proceeded. This shift was more pronounced in the 23-hour recovery simulation than in the 2.5-hour recovery simulation, reflecting a greater recovery of the cell population's capacity to tolerate additional radiation-induced damage. These differences in the damage-state distributions at the start of the second irradiation underlie the reappearance of the shoulder-like response seen in Fig. 3.

Thus, the model predicts the cell-level dynamics underlying the macroscopic increase in survival after an inter-fraction recovery interval.

*4.4 The quasi-threshold dose ($D_q$) and its biological significance*

The quasi-threshold dose $D_q$ is an index of shoulder width. For a split-dose curve based on a total dose of $D$, let $d = D - D_1$ denote the fractional dose delivered after the first dose ($D_1$; 5.05 Gy in this case) and $S_{rel}(d) = S(D_1 + d)/S(D_1)$ denote relative survival. The value of $D_q$ is the value of $d$ at which the extrapolated high-dose straight-line portion of the semilogarithmic survival curve intersects $S_{rel}$ = 1. Equivalently, on the absolute-survival plot, the extrapolated line intersects $S(D_1)$, rather than S = 1. This re-centered definition is used for the split-dose conditions in Table 2 and Fig. 6 (Okumura et al., 1974), and the single-dose definition and derivation are given in Appendix A.

$$D_q = \frac{\ln n}{\kappa} \tag{5}$$

For the 0-hour curve, $D_q$ was calculated using the conventional single-dose coordinate. For the 2.5-hour and 23-hour curves, dose was re-centered at 5.05 Gy and survival was normalized to $S$(5.05 Gy) before extrapolation. The resulting $D_q$ values are shown in Table 2.

The $D_q$ value increased with recovery time, and after 23 h of recovery, it approached the value produced under single-dose irradiation. This increase in $D_q$ indicates that the shoulder-like response to the second irradiation is greater after a longer recovery interval.

Although $D_q$ can be calculated separately for each recovery period treatment, it remains a phenomenological measure. Consequently, continuous changes in $D_q$ with changing recovery times cannot, by themselves, reveal the associated changes in cellular damage states. The present model helps bridge this gap by relating the macroscopic shoulder width represented by $D_q$ to the modeled distribution of damage states within the cell population.

### *4.5 Mechanism of $D_q$ recovery: Dynamics of the effective number of unhit targets*

In the framework of this model, the formation of the shoulder is closely related to the number of unhit targets present in the cell. Following the first irradiation, many cells die, and surviving cells receive multiple hits, accumulating sub-lethal damage. In this state, the cell population as a whole is vulnerable to subsequent hits, and additional small doses readily lead to lethality. This corresponds to a state in which the shoulder of the survival curve at the start of the second irradiation is small (small $D_q$).

During recovery, the damage distribution of the cells shifts toward a less-damaged state, i.e., the number of unhit targets increases. We quantified this shift using the "effective number of unhit targets," defined as the expected number of unhit targets (n – i) per cell in the original population. The lethal state contributes zero, so this quantity is not normalized by the surviving fraction:

$$\sum_{i=0}^{n-1} p^{(i)}(t) \times (n-i). \tag{6}$$

Through this approach, we linked the phenomenological parameter $D_q$ to dynamic variables within the cell population, a significant conceptual advance provided by this model. In conventional radiobiology, $D_q$ was a descriptive quantity derived from the shape of the curve, but this model allows for a mechanistic explanation of why $D_q$ recovery occurs: due to the restoration of the pool of unhit targets available to the cell population.

### *4.6. Exploratory model with a time-dependent recovery rate $\rho(t)$*

To examine the non-monotonic dependence of survival on the recovery interval, we fitted a model employing an exploratory extension in which $\rho(t)$ was allowed to vary with time during the recovery interval. The parameters $A_\rho$, $t_{c,\rho}$, $\sigma_\rho$, $\kappa_{base}$, and $\rho_{base}$ were determined by fitting this extended model to the recovery-time survival data. In this extension, $\rho(t)$ was modeled as a transient reduction from a baseline recovery rate (Eq. (4)).

As a result, the parameters were calculated (Table 3), and the experimental data were plotted alongside the extended model curve (Fig. 7). This analysis was performed separately from the main split-dose survival-curve fitting shown in Fig. 3.

The parameters derived in this section were obtained simultaneously from the Elkind–Sutton data for treatments receiving two radiation doses, in which cells experienced both radiation damage and cellular SLD recovery. Therefore, it is difficult to discuss each parameter’s relationship with biological actions separately, as discussed in Section 5.3. Whether this extended model provides a valid solution requires further investigation.

## 5. Discussion

### *5.1.Reproduction of the SLD recovery phenomenon by the model and its significance*

The mathematical model applied in this study describes the temporal evolution of damage accumulation and recovery within the classical multi-target, single-hit framework. It successfully reproduced the landmark split-dose dataset from Chinese hamster V79 cells reported by Elkind and Sutton (1960). As shown in Fig. 3, using a single parameter set including the number of targets ($n$), intrinsic radiosensitivity ($\kappa$), and recovery rate constant ($\rho$), the model consistently described not only the survival curve for the treatment without an intervening recovery interval but also the reappearance of a shoulder-like response after recovery intervals of 2.5 and 23 hours. This result suggests that the model provides an effective theoretical framework for quantitatively capturing the dynamic process of SLD recovery.

To clarify the significance of this approach, we compare the proposed model with major existing mathematical models used in radiobiology, particularly the linear-quadratic (LQ) model and classical target theory. This comparison is qualitative and conceptual, rather than a claim of superior predictive performance. The aim is not to show that the proposed model fits survival curves better than the LQ model, but to clarify the level at which each model represents the relevant biological processes.

#### *5.1.1. Time dependence in existing models and its limitations*

Although our model introduces the recovery term $\rho$, it inherits the zero-initial-slope property of the classical multi-target, single-hit model. For an initially undamaged population with $n > 1$, the initial slope of the single-dose survival curve is zero. This does not generally imply a zero initial slope for a re-centered split-dose curve, because some surviving cells may remain in damaged states when the second irradiation begins. The model may therefore overestimate survival in the very-low-dose region for an initially undamaged population where single-track lethal events may contribute. During acute high-dose-rate irradiation, the model approaches Eq. (3), so this structural limitation remains.

The LQ model uses coefficients $\alpha$ and $\beta$ for its linear and quadratic dose terms. These LQ coefficients are not analogous to the parameters $\kappa$ and $\rho$ in the present model: $\kappa$ is the damage-transition probability coefficient per unit dose, while $\rho$ is the first-order recovery-transition rate per unit time. Therefore, neither the notation nor the biological interpretation are interchangeable. The LQ model has a finite initial slope and can be extended with the Lea–Catcheside time factor, whereas the present model explicitly follows damage-state probabilities through time (Brenner, 2008).

While this Lea–Catcheside time factor-corrected LQ model is highly useful in clinical applications, its approach has conceptual limitations. The LQ model is essentially a static model that describes the final "result" (survival in the current context) after a specific dose is given, and the Lea–Catcheside time factor functions as a supplementary "correction" to account for time effects within that static framework. In contrast, the approach adopted in this study differs fundamentally in its modeling philosophy. Our model is based on physically intuitive target theory and explicitly describes the two competing "processes"—damage (hit) induction and biological repair—as a system of simultaneous ordinary differential equations with time as an independent variable, creating an inherently dynamic framework that simulates how the state of the cell population transitions over time. The LQ model provides a compact phenomenological description of the final surviving fraction as a function of dose. In contrast, the present model explicitly describes the time evolution of damage-state probabilities during irradiation and recovery. Thus, the distinction between the two models primarily lies not in their goodness of fit to survival data, but in the level of representation: the LQ model summarizes the final dose–response relationship, whereas the present model resolves the processes by which damage is accumulated and repaired before the final surviving fraction is determined. This "visualization of the process" is one of the key contributions of this study.

A further distinction is the explicit treatment of dose rate. In the basic LQ formulation, dose-rate effects are not represented as an independent input variable; rather, they are effectively absorbed into fitted radiosensitivity parameters and may require additional time-correction factors. In the present ODE framework, the dose rate $\gamma$ is incorporated directly as the protocol-defined input that drives transitions from lower-damage to higher-damage states during irradiation, while $\rho$ represents biological recovery. This explicit separation between physical dose delivery and biological recovery improves model parameter interpretability and facilitates comparison among different irradiation protocols.

### *5.1.2. Positioning of this model in the trend of mechanistic models*

More sophisticated models aiming to understand the biophysical mechanisms underlying the actions of radiation in tissues have also been proposed. These models attempt to complement the phenomenological description of the LQ model and link molecular-level knowledge, such as the dynamics of DNA double-strand breaks, to cell survival. Representative mechanistic models, including the LPL model, RMF model, and MKM, have greatly contributed to the theoretical deepening of radiobiology. However, due to their high levels of sophistication, they require many parameters and are mathematically complex, making intuitive understanding and the visualization of phenomena difficult.

Our model, thus, occupies a niche of simplicity and interpretability in this theoretical landscape. While it is mechanistic in the sense that it explicitly represents damage induction and biological recovery as state transitions, the model is characterized by a small number of interpretable parameters: the number of targets ($n$), the dose-related damage transition coefficient ($\kappa$), the dose-rate input ($\gamma$), and the recovery rate ($\rho$). It is mathematically tractable and can directly simulate the temporal evolution of the damage-state distribution $p^{(i)}(t)$ within the cell population and readily allows visualization. Therefore, this model provides a useful framework for quantitatively reproducing SLD recovery while facilitating a process-level interpretation of the underlying cell population dynamics.

*5.2.Mechanistic insights into the reappearance of the shoulder-like response*

One of the characteristic aspects of fractionated irradiation reproduced by this model is the shoulder-like response appearing after an inter-fraction recovery interval. This response reflects the cell population's capacity to tolerate and recover from accumulated sub-lethal damage.

The effective number of unhit targets approaches an asymptotic recovery level, defined here as the limiting value predicted in the relevant recovery-only system as recovery time tends to infinity. In the fitted single-rate model, the time to reach 95% of this total model-predicted change is approximately 8.3 h; however, note that this is a model-derived summary, rather than a sharply defined biological threshold.

This correspondence should be interpreted cautiously. Double-strand break repair in mammalian cells is primarily carried out via two major pathways. One is the rapid but error-prone non-homologous end joining (NHEJ), which has a half-life of tens of minutes. The other is the more time-consuming but accurate homologous recombination (HR), which has a half-life extending to several hours (Ahnstrom et al., 2000). The SLD recovery observed in Elkind and Sutton's experiment is the macroscopic result of these asynchronous repair pathways. The model-derived 8.3-hour timescale may therefore reflect the population-wide average recovery of effective damage tolerance, rather than the effects of a single molecular repair pathway. Thus, we use this value as a phenomenological guide to the timescale of near-complete recovery, not a precise biological threshold.

*5.3.Biological plausibility of model parameters*

Verifying whether the parameters determined in this study ($n = 5, \kappa = 0.7\ 1/\mathrm{Gy}, \rho = 0.006\ 1/\mathrm{min}$; see Table 1) are not merely fitting parameters but have relevance to biological entities is a crucial part of validating the model.

*5.3.1. Interpretation of the recovery rate constant ρ*

The fitted recovery rate was $\rho = 0.006$ 1/min. For recovery of an individual effective damage event, the half-life is $t_{1/2} = \ln(2)/\rho \approx 116$ min (~120 min). This is the per-event recovery timescale. The transition from state i to state i − 1 occurs at rate iρ, so its waiting-time half-life is ln(2)/(iρ). It is not the time required for a cell to traverse all damaged states back to $p^{(0)}$.

Because the experiment measures clonogenic survival in an asynchronous population, $\rho$ should be interpreted as an effective population average rather than the rate of a single molecular pathway. Population-level observations conflate cell-cycle states, repair-pathway availabilities, and lesion complexities; NHEJ and HR, thus, contribute to the fitted macroscopic timescale in aggregate (Shibata et al., 2011; Schipler and Iliakis, 2013).

Therefore, the single recovery rate constant in this model, $\rho$, does not directly reflect the rate constant of a specific molecular reaction. Rather, it is appropriate to interpret it as an "effective" or "population-averaged" rate constant that describes the phenomenological process of recovery from SLD observed across the entire cell population, where diverse cell states and damage types coexist. The calculated time of 120 minutes represents a timescale over which the entire process—from the initial stage dominated by rapid NHEJ repair to the later stages in which HR handles more complex damage types—is approximated by a single exponential function. This interpretation suggests that, while admitting the model is a simplification, its parameters capture the average behavior of biological reality.

### *5.3.2. Reconsideration of the biological entities "hits" and "targets"*

As an exploratory biological interpretation, we examined whether the fitted target number, $n = 5$, could be broadly related to genome-scale radiosensitive regions. This calculation should not be interpreted as associating individual model hits with individual physical DSBs. Instead, the target number is treated as an effective coarse-grained quantity designed to reflect the subset of radiation-induced lesions, including but not limited to DSBs, that contribute to clonogenic inactivation. Based on the mean inactivation dose M calculated using Eq. (B.8) and reported DSB induction yields per Gy as a reference scale, we estimated the approximate fraction of genomic regions not represented by the effective target number. This exploratory calculation gave a value broadly comparable to that of non-coding DNA fractions reported in Chinese hamster-derived cells, but this result should be regarded only as a macroscopic consistency check rather than a molecular identification of the model targets.

Indeed, this estimation requires careful re-evaluation when compared with more detailed biological knowledge. First, DSB induction varies with cell type and experimental conditions. The DSB yields induced by low-LET radiation in mammalian cells are generally reported to be in the range of 20–40 per Gy per cell (Lomax et al., 2013). In V79 cells and related CHO cells, reported values vary; for instance, one study reported an increase of about 19.7 $\gamma$-H2AX foci (a surrogate marker for DSBs) per Gy in V79 cells (Kotenko et al., 2013), while another reported a value of 52.58 per Gy in CHO-K1 cells (Mori et al., 2018). For an illustrative DSB yield Y ranging from 20 to 55 per Gy per cell, we estimated the fraction not represented by the effective targets as $1 - n/(Y\ M)$. Using $n = 5$ and $\kappa = 0.716$ 1/Gy from Table 1 gives $M \approx 3.189$ Gy and estimated proportions of 92.2–97.1%. These are heuristic estimates, not direct measurements of non-coding DNA, and depend strongly on the assumed relationship between effective targets and DSB yields.

Second, the assumption that the model parameter $n = 5$ accurately represents the number of lethal DSBs in coding regions across a vast genome is a very bold simplification. In actual cells, DSB inducibility and repair efficiency are heterogeneous, depending on genomic location, and not all coding regions are uniformly involved in cell death. Therefore, while the correlation based on the estimated non-coding DNA proportion described above is an interesting result, it should be regarded as suggesting a correlation at a macroscopic level, and further verification of the biological entities of the "targets" in this model is needed.

### *5.4. Model limitations and future extensibility*

The mathematical model applied in this study achieved significant success in capturing the basic dynamics of SLD recovery, but several limitations stemming from its simplified assumptions are also apparent. Recognizing these limitations and discussing directions for future improvement is essential for further developing this research.

#### *5.4.1. Inheritance of the Zero-Slope Problem in the Low-Dose Region*

Although the mathematical model applied in this study describes dynamic recovery processes through the introduction of the recovery term $\rho$, it inherits a theoretical limitation of the classical multi-target, single-hit model on which it is based: its zero-initial-slope property. For target numbers greater than one, the initial slope of the survival curve becomes zero as $D$ approaches zero. Thus, the model can overestimate survival at very low doses, where single-track lethal events or other low-dose effects may contribute to cell killing. As discussed in Sections 4.1 and 4.2, during high-dose-rate acute irradiation, under which the contribution of $\rho$ is negligible, our model is effectively simplified to the classical multi-target, single-hit model (Eq. (3)) and therefore retains this structural property.

This limitation clarifies the present model's intended scope. The model is not intended to replace the LQ model as an empirical description of the low-dose region. Rather, its purpose is to provide a simple and interpretable dynamic framework for describing and visualizing the time-dependent process of SLD recovery after fractionated irradiation. With this aim, it prioritizes the explicit representation of damage-state transitions and recovery dynamics over empirical flexibility in the low-dose region.

This zero-initial-slope property can also affect the structure of residuals in the least-squares fitting. Although we used logarithmic residuals to reduce the imbalance between high- and low-survival regions, the model's residual errors are not expected to be perfectly independent and identically distributed across dose and recovery conditions. In particular, the structural tendency of the multi-target model to overestimate survival at very low doses can lead the fitting procedure to balance errors across the full dataset, occasionally resulting in minor underpredictions at higher doses or when applied to 0-hour recovery treatment data.

A possible extension of the model would be to introduce an additional single-hit lethal component, analogous to the linear term in the LQ model. Such an extension could improve the description of survival in the low-dose region and reduce the zero-initial-slope limitation. However, it would also introduce additional parameters, reducing the simplicity and interpretability of the present Markov-chain-based formulation. Therefore, in this study, we preserved the current parameter structure and explicitly acknowledge this issue as an inherent limitation of applying classical target theory across all dose ranges.

### *5.4.2 Impact of cell proliferation in long-term recovery*

Detailed observation of Fig. 3 reveals a tendency for the model's predictions to slightly underestimate experimental survival values, with a larger discrepancy seen for the data from the treatment with the long, 23-hour recovery period. The most likely cause of this discrepancy is the effect of cell proliferation (repopulation) during the recovery period.

The cell cycle of Chinese hamster V79 cells is generally reported to be about 10–14 hours, depending on culture conditions (Klevecz, 1976). Therefore, a 23-hour recovery period is sufficient for cells surviving the first irradiation to divide at least once, increasing the cell number. In Elkind and Sutton (1960), cells in the logarithmic growth phase were used, and there is no clear description of the use of special measures to suppress cell number increase during this long recovery interval. Therefore, it is highly probable that the increase in survival observed after 23 hours in the experiment includes an added effect of increases in total cell numbers due to cell division in addition to the recovery of individual cells' viabilities due to SLD recovery. Since the mathematical model used in this study assumes a constant cell number and considers only radiation damage and repair processes, it cannot describe this proliferation effect. We believe this to be the main cause of the fitting discrepancy seen when using the 23-hour recovery data.

To address this limitation, future model improvements could involve introducing a cell proliferation term into the differential equation system. The simplest approach would be to assume that surviving cells (cells from $p^{(0)}$ to $p^{(n-1)}$) proliferate exponentially or logistically at a certain rate during the recovery period and incorporate this effect into the model. A more sophisticated approach would be to extend this approach to a multi-state model that tracks the number of cells in each phase of the cell cycle as separate variables. Such extensions could significantly improve the predictive accuracy and applicability of the model when examining longer-term cellular responses or the combined effects of cell cycle-specific agents (Richard et al., 2007).

### *5.5. Cell cycle synchronization and variation in radiosensitivity*

Another important limitation of this model is that the recovery rate constant $\rho$ was assumed to be unchanging over time. While this assumption was effective for describing average behavior across specific recovery periods, like 2.5 or 23 hours, it cannot capture the dynamic changes of the cell population during the recovery time.

In Elkind and Sutton (1960), in addition to the dataset we attempted to reproduce in this paper, there is data showing final survival fractions under varying recovery periods (black dots in Fig. 7). Note that the horizontal axis shows the recovery time between the first and second irradiations, and the vertical axis represents cell survival after the second irradiation. Looking at this data, it is clear that survival does not increase monotonically with recovery time; instead, a "survival reversal phenomenon" is observed in which survival temporarily decreases at a certain time. Such complex patterns, with survival fluctuating across the repair period, particularly the phenomenon in which a temporary decrease (dip) is seen, have also been reported in fractionated irradiation experiments using other cell lines (Rao et al., 2007).

This phenomenon cannot be explained by the current model, given that $\rho$ remains constant. The underlying mechanism is thought to be strongly related to the cell cycle and associated variations in radiosensitivity. This phenomenon is generally known as "redistribution," one of the 4 Rs of radiobiology (Zaider et al., 1996). That is, the cell population surviving the first irradiation proceeds through the cell cycle in a somewhat synchronized state. Cells have significantly different radiosensitivity depending on the phase of the cycle. Generally, late G1 to S phase cells show resistance, while G2/M phase cells are most sensitive (Sinclair and Morton, 1966). The first irradiation selects against cells in sensitive phases, leaving mainly cells in resistant phases, synchronizing the cell population. Subsequently, when this synchronized population transitions collectively over time to the next, more sensitive phase (e.g., G2/M phase), the apparent radiosensitivity of the entire population temporarily increases. If the second irradiation is performed at this timing, the lethal effect is enhanced. A constant $\rho$ value cannot capture such time-dependent fluctuations in sensitivity. To address this problem, an extension in which the recovery rate $\rho$ is not regarded as constant but treated as a time-dependent function $\rho(t)$ that varies with recovery time $t$ could be considered. This extension proposal is discussed in Section 3.3.

*5.6.Contributions of this study to theoretical and radiation biology*

Our multi-target, single-hit model considering the recovery process, applied and validated in this study, succeeded in uniformly describing the time-dependent phenomenon of recovery from SLD, a classic problem in radiobiology, within a mathematical framework. From the perspective of theoretical biology, this demonstrates the effectiveness of modeling the heterogeneous damage states of cell populations and stochastic recovery processes as a system of differential equations and analyzing their dynamic behavior. In particular, tracking the proportion by hit numbers $p^{(i)}(t)$ within the cell population as a time-dependent variable allowed us to visualize not only the prediction of final cell survival but also the state transition dynamics within the cell population leading up to it, which offers new insights for understanding the complex processes of radiation response.

In the field of radiobiology, this study is considered to contribute in three ways:

First, the model supports a quantitative evaluation of SLD recovery. Recovery capacity can be compared across conditions using the recovery-rate constant $\rho$, the derived repair half-life, and changes in the effective quasi-threshold dose $D_q$ with recovery time. These quantities provide a more explicit description of SLD recovery than purely phenomenological comparisons of survival curves.

Second, the model provides effective parameters ($n$, $\kappa$, and $\rho$) that summarize cellular radiosensitivity and recovery capacity. In this study, their possible relationships to biological quantities were explored using measures like repair half-life and non-coding DNA proportions. Applying the model to additional cell types and experimental conditions may help determine whether changes in these effective parameters are associated with specific molecular or cellular mechanisms.

Third, the model may be useful in the pre-clinical investigation of time-dependent radiation responses. We do not intend to suggest that the present model can replace the LQ model in clinical treatment planning. The LQ model remains compact, clinically established, and practical for treatment planning. By contrast, the present model is intended to provide a process-oriented framework for visualizing damage-state distributions and recovery dynamics after irradiation. After further validation using additional cell types, irradiation protocols, and clinically relevant datasets, such a framework may help generate hypotheses about fractionation intervals, follow-up treatment timing, or the use of repair inhibitors in pre-clinical studies.

Although this model has limitations, such as the influence of cell proliferation and cell cycle and the complexity of recovery rates, as mentioned in Section 5, incorporating these elements in future model formulations holds the potential to describe and predict even broader radiation biological phenomena. While this model does not resolve the fundamental issue of the classical target theory in which the initial slope of the survival curve becomes zero in the low-dose region, the main contribution of this study is that it quantitatively described the entire process of SLD recovery and provided a dynamic framework for visualizing the underlying cell population dynamics. This offers a valuable process-oriented perspective that complements traditional static result-descriptive models like the LQ model. Overall, this study is expected to be an important step towards deepening the theoretical understanding of the dynamics of radiation-induced cell death and connecting that knowledge to practical applications such as radiotherapy and protection.

## 6. Conclusion

In this study, we quantitatively described sub-lethal damage (SLD) recovery by analyzing the classical split-dose experimental data of Elkind and Sutton using a mechanistic mathematical model that extends the classical multi-target, single-hit framework with explicit recovery transitions among damage states. Using a common parameter set consisting of the number of targets ($n$), intrinsic radiosensitivity ($\kappa$), and recovery rate constant ($\rho$), the model reproduced the major features of the split-dose survival curves for treatments with recovery intervals of 0, 2.5, and 23 hours.

The significance of this study extends beyond reproducing final survival fractions. The present model represents damage accumulation and biological recovery as state transitions in a Markov-chain-based system of ordinary differential equations. It therefore visualizes the redistribution of damage states across the cell population during irradiation and recovery. This process-level representation links macroscopic quantities, such as the quasi-threshold dose $D_q$ and the reappearance of a shoulder after an inter-fraction interval, to changes in the modeled damage-state distribution.

Furthermore, the recovery half-life estimated from the fitted recovery rate was approximately 120 min, which is broadly consistent with reported timescales of DNA double-strand break repair in mammalian cells. This agreement suggests that the fitted parameters can be interpreted as effective, population-averaged quantities that link macroscopic survival recovery to underlying biological recovery processes.

The present formulation also has important limitations. The main model assumes a single effective recovery rate and does not explicitly incorporate cell proliferation, cell-cycle redistribution, or multiple repair pathways. These factors may become important during long recovery intervals, such as that of the 23-hour condition. In addition, the model inherits the zero-initial-slope property of the classical multi-target, single-hit model and is therefore not intended to replace empirical low-dose descriptions, such as the LQ model. Future extensions incorporating proliferation, cell-cycle structure, and additional lethal components will be necessary to evaluate the applicability of the framework under broader experimental conditions.

In conclusion, the proposed model provides a simple, interpretable dynamic framework for connecting radiation-induced damage accumulation, biological recovery, and clonogenic survival. By visualizing damage-state transitions, the model offers a process-oriented perspective that complements conventional results-descriptive models. Further application to different cell types, radiation qualities, dose rates, and fractionation schedules will be necessary to assess the generality of this framework and to develop it into a more broadly predictive model of radiation-induced cell death.

**Author Contributions : CRediT**

Renta Shibata: Investigation, Visualization, Writing – original draft.
Fuminobu Sato: Supervision.
Yoko Akiyama: Supervision.
Yuichiro Manabe: Methodology, Validation, Writing – review & editing.

**Funding Sources**

This work was supported by JSPS KAKENHI Grant Number 22K04987.

**Declaration of generative AI and AI-assisted technologies in the writing process**

During the preparation of this work, the authors used Google Gemini and OpenAI Codex to improve the readability and language of the manuscript, assist with translation and formatting, and support the preparation of revised figures and data presentation. After using these tools/services, the authors reviewed and edited the content as needed and take full responsibility for the content of the published article.

**Declaration of competing interests**

The authors declare that they have no known competing financial interests or personal relationships that could have appeared to influence the work reported in this paper.

## 7. Appendix

### A. Derivation of the quasi-threshold dose ($D_q$)

Let S(D) denote the surviving fraction in the multi-target, single-hit model, where $D$ is the dose, $n$ is the number of targets, and $\kappa$ is the intrinsic radiosensitivity:

$$S(D) = 1 - (1 - e^{-\kappa D})^n \qquad (A.1)$$

When $D$ is sufficiently large, $\exp(-\kappa D)$ is small, and a first-order binomial expansion then gives Eq. (A.2).

$$S(D) \approx n \cdot e^{-\kappa D} \qquad (A.2)$$

This approximation describes the straight-line portion of the survival curve beyond the shoulder. Taking the natural logarithm of both sides gives

$$\ln S(D) \approx \ln n - \kappa D \qquad (A.3)$$

Equation (A.3) is a straight line on a semilogarithmic plot, with slope $-\kappa$ and intercept $\ln(n)$. The quasi-threshold dose $D_q$ is defined by extrapolating this line to $S(D) = 1$. Therefore, setting $\ln[S(D)] = 0$ and $D = D_q$ in Eq. (A.3) gives

$$0 = \ln n - \kappa D_q$$

And solving for $D_q$ gives

$$D_q = \frac{\ln n}{\kappa} \qquad (A.4)$$

Thus, the quasi-threshold dose is $D_q = \ln(n)/\kappa$.

**B. The high-dose slope parameter and mean inactivation dose**

In target theory, the mean lethal dose $D_0$ is the dose associated with one lethal event per target on average.

The following derivation follows the standard target theory formulation. In the single-target, single-hit model, a dose $D_0$ produces one hit on average, and a dose $D$, therefore, produces $D/D_0$ hits on average:

$$S(t) = e^{-\frac{D}{D_0}} \qquad (B.1)$$

Taking the natural logarithm of both sides of Eq. (B.1):

$$lnS(t) = -\frac{D}{D_0} \quad (B.2)$$

Equation (B.2) is a straight line with a slope $-1/D_0$ when survival is plotted on a logarithmic scale against dose. At $D = D_0$, the surviving fraction is exp(−1) = 1/e.

For the *n*-target, single-hit model, the surviving fraction is given by Eq. (3). In the high-dose region, where $D$ is much larger than $D_0$, Eq. (3) can be approximated by Eq. (B.3).

$$S(t) = 1 - (1 - e^{-\frac{D}{D_0}})^n \approx ne^{-\frac{D}{D_0}} \quad (B.3)$$

Taking the natural logarithm of both sides of Eq. (B.3) leads to Eq. (B.4):

$$lnS(t) = \ln(n) - \frac{D}{D_0} \quad (B.4)$$

On a semilogarithmic plot, Eq. (B.4) approaches a straight line with a slope of $-1/D_0$, as in the single-target, single-hit model. Thus, $D_0 = 1/\kappa$ is the high-dose slope parameter. We distinguish this quantity from the mean inactivation dose M, defined as the mean of the cell-inactivation dose distribution derived from the full survival curve.

Consider an increase in absorbed dose from $D$ to $D + dD$. Let S($D$) denote the surviving fraction in the *n*-target single-hit model, as given by

$$S(D) = 1 - (1 - e^{-\kappa D})^n \quad (B.5)$$

Differentiating S($D$) with respect to $D$ gives

$$\frac{dS}{dD} = -n\kappa e^{-\kappa D}(1 - e^{-\kappa D})^{n-1} \quad (B.6)$$

This derivative is the change in surviving fraction per unit dose. Because $D$ is measured in Gy, d$S$/d$D$ has units of 1/Gy, and since S($D$) decreases as dose increases, dS/d$D$ is negative.

Therefore, the infinitesimal fraction of cells that die between $D$ and $D + dD$ is:

$$d[1 - S] = -dS = -(dS/dD)dD \quad (B.7)$$

Equivalently, the dose density of cell death is

$$d[1 - S]/dD = -dS/dD.$$

The fraction of cells that die between $D$ and $D + dD$ is $d[1 - S]$, so the contribution of this interval to the dose-weighted mean is *D* *d*[1 − *S*]. Integrating over all doses gives the mean inactivation dose M:

$$M = \int_0^\infty n\kappa D e^{-\kappa D}(1 - e^{-\kappa D})^{n-1} dD$$
$$= n\kappa \int_0^\infty D e^{-\kappa D}(1 - e^{-\kappa D})^{n-1} dD \qquad (B.8)$$

For n = 5 and the rounded value κ = 0.7 1/Gy, Eq. (B.8) gives M ≈ 3.262 Gy, whereas $D_0$ ≈ 1.429 Gy. Using κ = 0.716 1/Gy from Table 1 gives M ≈ 3.189 Gy. The two dose measures coincide only for n = 1, for which Eq. (B.8) reduces to

$$M = \kappa \int_0^\infty D e^{-\kappa D} dD$$
$$= \frac{1}{\kappa} \qquad (B.9)$$

Substituting $D_0 = 1/\kappa$ and $n = 1$ into Eq. (B.5) gives Eq. (B.10), which is consistent with Eq. (B.4):

$$S(D_0) = \frac{1}{e} \qquad (B.10)$$

**C. Captions**

**Fig. 1. Experimental survival data, derived from Fig. 11 of Elkind and Sutton (1960).** Data points represent the fraction of Chinese hamster V79 cells surviving exposure to fractionated X-ray irradiation. Black circles denote the nonfractionated (0-hour) condition data, blue open squares the 2.5-hour interval data, and red filled triangles the 23-hour interval data. The lower axis gives total dose; the upper axis gives the second dose administered for the split-dose series, after a first 5.05 Gy exposure.

These data points were created by digitizing and extracting the points in Fig. 11 of Elkind and Sutton (1960) using the graph digitizing system (GSYS) for the purpose of parameter estimation in this study.

**Fig. 2. Transition scheme for the conventional damage-only model and the proposed dynamic model.** Forward transitions are driven by $(n - i)\kappa\gamma(t)$, while recovery transitions are driven by $i\rho(t)$. The lethal state $p_n$ is absorbing: cells cannot recover from this state. The variables $\kappa$ and $\rho$ are distinct from the $\alpha$ and $\beta$ coefficients found in LQ models.

**Fig. 3. Model predictions and experimental data from Elkind and Sutton (1960).** Data points (symbols) and model curves (solid lines) are color coded for the 0-hour, 2.5-hour, and 23-hour recovery period treatment conditions. Split-dose curves and data are shown only from a total dose of 5.05 Gy (vertical dashed line) onward. The lower axis gives total dose, and the upper axis gives second dose. The endpoint is immediately after the second irradiation.

**Fig. 4. Local sensitivity analysis of the approximations of the model not including recovery during acute irradiation.** The solid black line represents the survival curve calculated using the model without recovery during irradiation, whereas the dashed red line represents the calculation produced by the model including simultaneous recovery during irradiation. Under the short high-dose-rate irradiation condition examined here, the difference between the two calculations was very small. This comparison was used only to assess the local effect of recovery during the short dose-delivery period.

**Fig. 5. Simulated temporal evolution of intracellular damage states under representative split-dose protocols: (a) 0-hour recovery condition, (b) 2.5-hour recovery condition, and (c) 23-hour recovery condition.** Each panel shows the time evolution of the cell fractions $p_i(t)$ for damage states $i$ = 0, ..., $n$ during the first irradiation, recovery interval, and second irradiation. To facilitate comparison among protocols with very different recovery durations, each panel is displayed using a segmented time axis with the same three phases. The actual duration of each irradiation phase is indicated in the x-axis labels: the 5.05 Gy irradiation at 7.22 Gy/min lasted ~0.70 min. The recovery intervals were 0, 150, and 1380 min for the 0-hour, 2.5-hour, and 23-hour conditions, respectively. The simulation endpoint is defined as the end of the irradiation–recovery–irradiation protocol, immediately after the second irradiation, not as the final colony-scoring time in the original experiment.

**Fig. 6. Definition of $D_q$. For split-dose curves, $d = D - 5.05$ Gy and $S_{\mathrm{rel}}(d) = S(5.05\ \mathrm{Gy} + d)/S(5.05\ \mathrm{Gy})$.** The high-dose straight-line extrapolation intersects $S_{\mathrm{rel}} = 1$, while, equivalently, the absolute-survival line intersects $S$(5.05 Gy).

**Fig. 7. Exploratory time-dependent-parameter extensions for recovery-time data derived from Fig. 16 of Elkind and Sutton (1960), specifically from the two-587-rad exposure series.** (a) Black circles: data; solid blue line: time-dependent $\rho(t)$ model; solid red line: alternative time-dependent $\kappa(t)$ model; dashed gray line: constant-parameter model. (b) $\rho(t) = \max[0, \rho_{\text{base}} - A_\rho \exp\{-(t - t_{c,\rho})^2/(2\sigma_\rho^2)\}]$ and its constant baseline. (c) $\kappa(t) = \kappa_{\text{base}} + A_\kappa \exp\{-(t - t_{c,\kappa})^2/(2\sigma_\kappa^2)\}$ and its constant baseline; the constant recovery rate used in that alternative model is also included. These extensions are exploratory and were not used for main parameter estimation.

# Tables

New submission related to JTB-D-25-01100R5

**Table 1. Estimated model parameters and fixed protocol inputs.**

| Symbol | Description | Unit | Value | Status |
|---|---|---|---|---|
| n | Number of effective targets | — | 5 | Estimated; rounded to integer |
| κ | Damage-transition coefficient | 1/Gy | 0.716 (reported as 0.7) | Estimated |
| ρ | Recovery-transition rate | 1/min | 0.00596 (reported as 0.006) | Estimated |
| γ | Dose rate during irradiation | Gy/min | 7.22 | Fixed protocol input |
| $D_1$ | First irradiation dose | Gy | 5.05 | Fixed protocol input |
| $T_1$ | First irradiation duration ($D_1/\gamma$) | min | 0.699 | Calculated |

**Table 2. Quasi-threshold dose after re-centering split-dose curves at 5.05 Gy.**

| Condition | $D_q$ (Gy) |
|---|---|
| 0 h (single/nonfractionated) | 2.24 |
| 2.5-h recovery | 1.37 |
| 23-h recovery | 2.17 |

**Table 3. Parameters used in the exploratory time-dependent extensions.**

| Model | Symbol | Description | Unit | Value |
|---|---|---|---|---|
| ρ(t) | $A_\rho$ | Amplitude of transient reduction | 1/min | 0.01054 |
| ρ(t) | $t_{c,\rho}$ | Center time | min | 365.33 |
| ρ(t) | $\sigma_\rho$ | Characteristic width | min | 185.39 |
| ρ(t) | $\kappa_{\text{base}}$ | Constant damage coefficient | 1/Gy | 0.81076 |
| ρ(t) | $\rho_{\text{base}}$ | Baseline recovery rate | 1/min | 0.012414 |
| κ(t) | $A_\kappa$ | Amplitude of transient increase | 1/Gy | 0.07548 |
| κ(t) | $t_{c,\kappa}$ | Center time | min | 284.73 |
| κ(t) | $\sigma_\kappa$ | Characteristic width | min | 102.21 |
| κ(t) | $\kappa_{\text{base}}$ | Baseline damage coefficient | 1/Gy | 0.78147 |
| κ(t) | $\rho_{\text{const}}$ | Constant recovery rate | 1/min | 0.017855 |

Figure 1

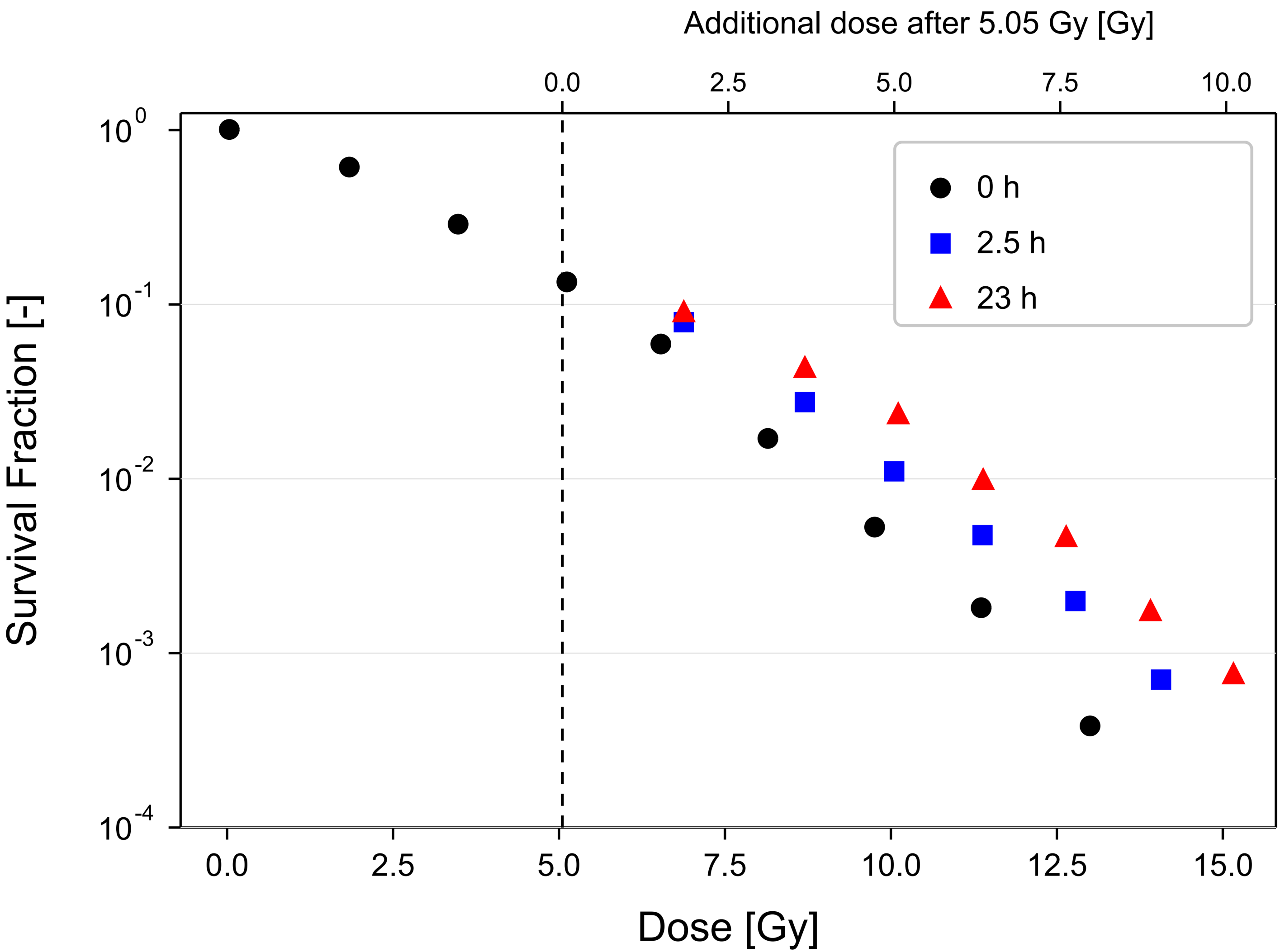

Additional dose after 5.05 Gy [Gy]
0.0
2.5
5.0
7.5
10.0
$10^0$
$10^{-1}$
$10^{-2}$
$10^{-3}$
$10^{-4}$
Survival Fraction [-]
0 h
2.5 h
23 h
0.0
2.5
5.0
7.5
10.0
12.5
15.0
Dose [Gy]

Figure 2

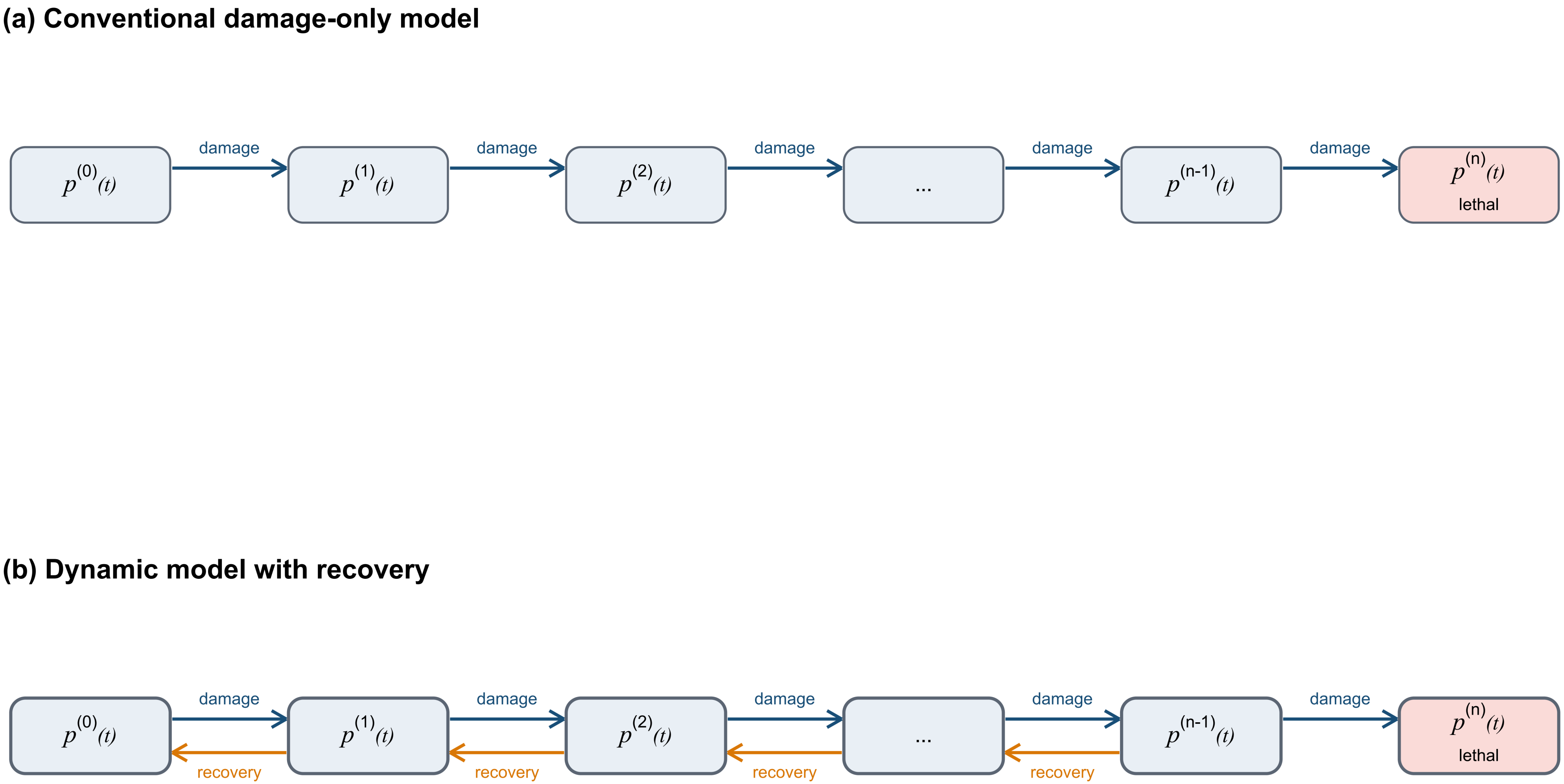

Damage transition rate: (n-i) $\kappa\gamma(t)$. Recovery transition rate: (i+1) $\rho(t)$.
The lethal state is absorbing; recovery occurs only among viable states.

Figure 3

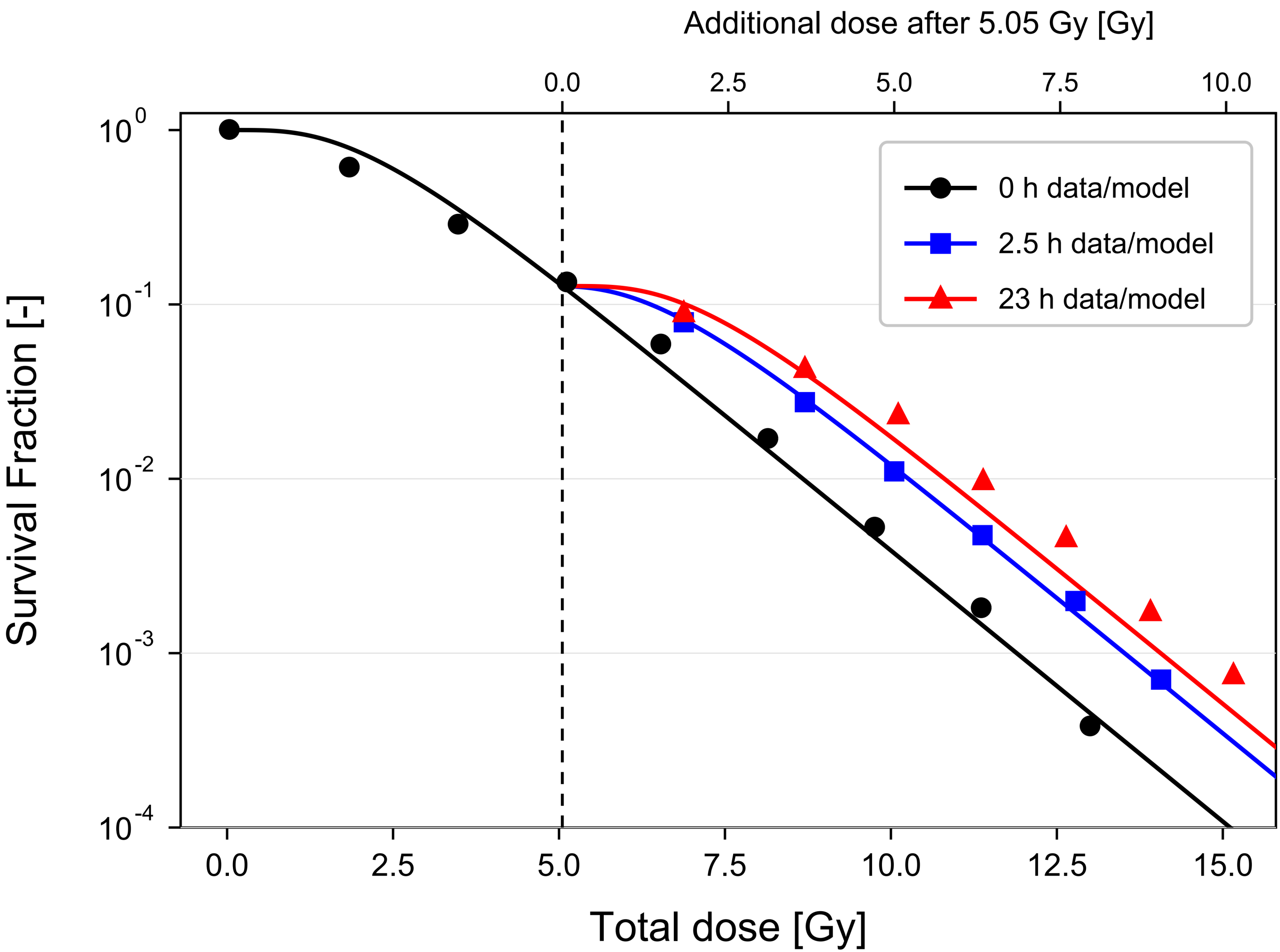
Additional dose after 5.05 Gy [Gy]
0.0
2.5
5.0
7.5
10.0
$10^{0}$
$10^{-1}$
$10^{-2}$
$10^{-3}$
$10^{-4}$
Survival Fraction [-]
0 h data/model
2.5 h data/model
23 h data/model
0.0
2.5
5.0
7.5
10.0
12.5
15.0
Total dose [Gy]

Figure 4

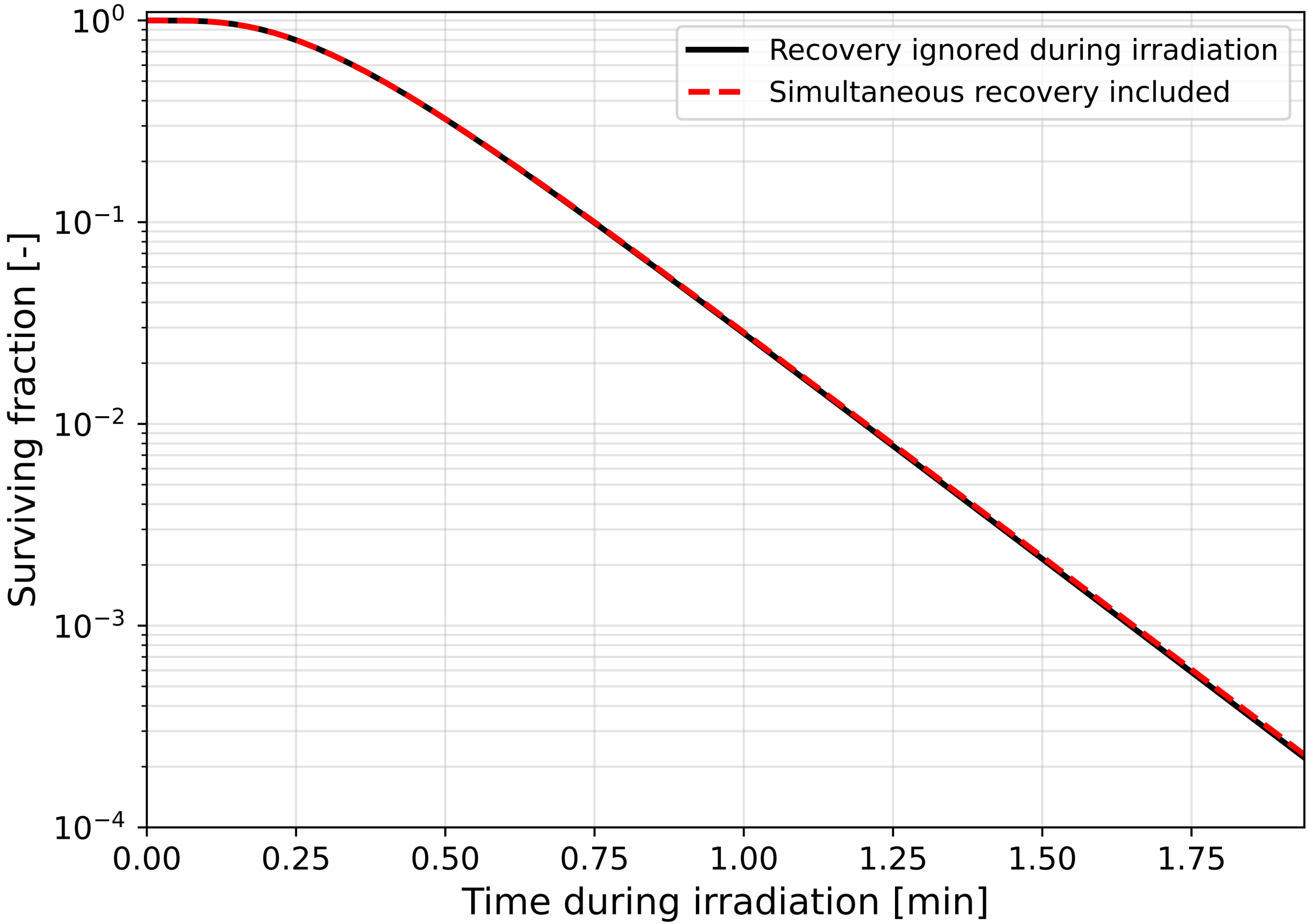

Recovery ignored during irradiation
Simultaneous recovery included
Surviving fraction [-]
Time during irradiation [min]
10^0
10^−1
10^−2
10^−3
10^−4
0.00
0.25
0.50
0.75
1.00
1.25
1.50
1.75

Figure 5

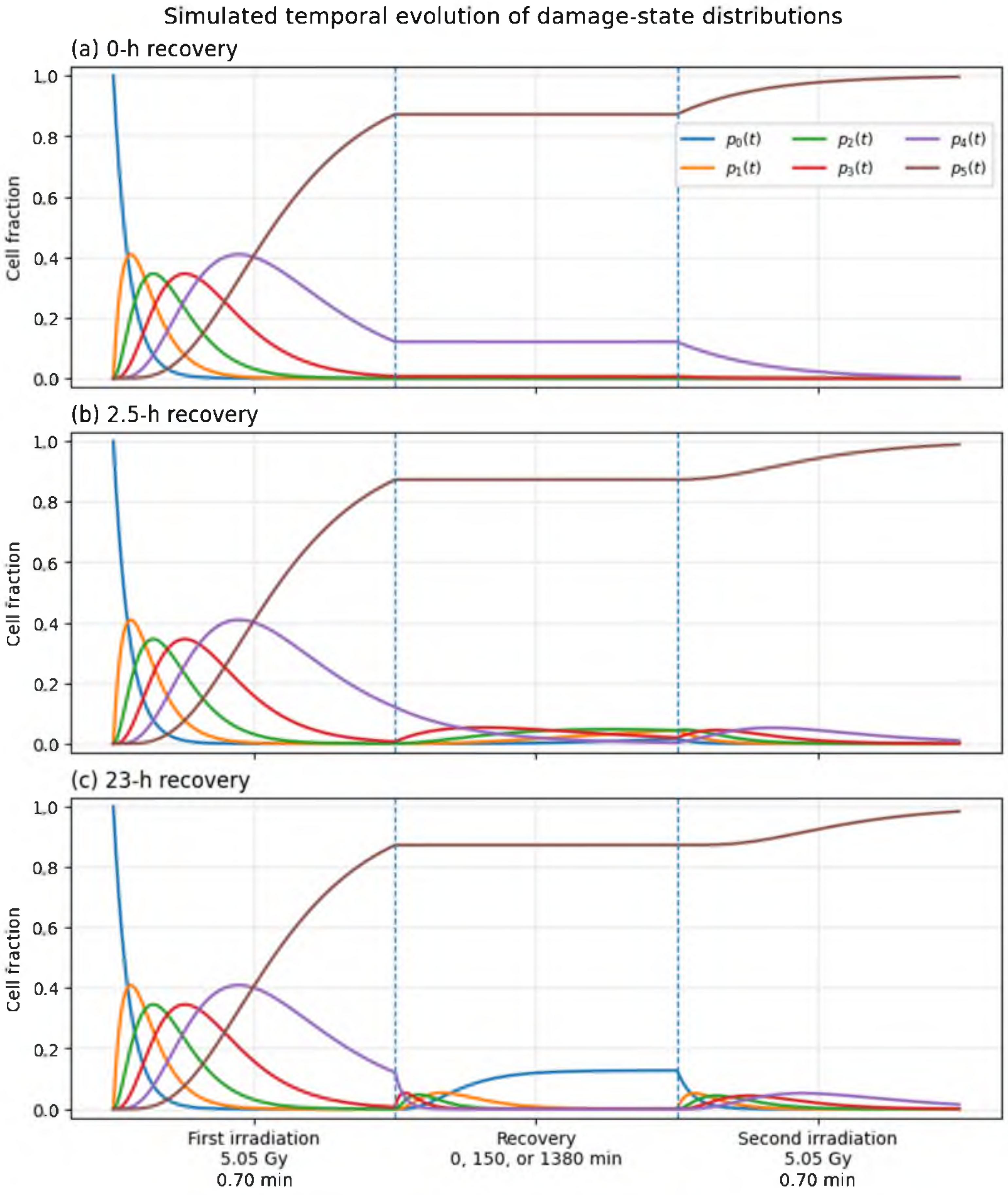

Simulated temporal evolution of damage-state distributions
(a) 0-h recovery
(b) 2.5-h recovery
(c) 23-h recovery
Cell fraction
1.0
0.8
0.6
0.4
0.2
0.0
$p_0(t)$
$p_1(t)$
$p_2(t)$
$p_3(t)$
$p_4(t)$
$p_5(t)$
First irradiation
5.05 Gy
0.70 min
Recovery
0, 150, or 1380 min
Second irradiation
5.05 Gy
0.70 min

Figure 6

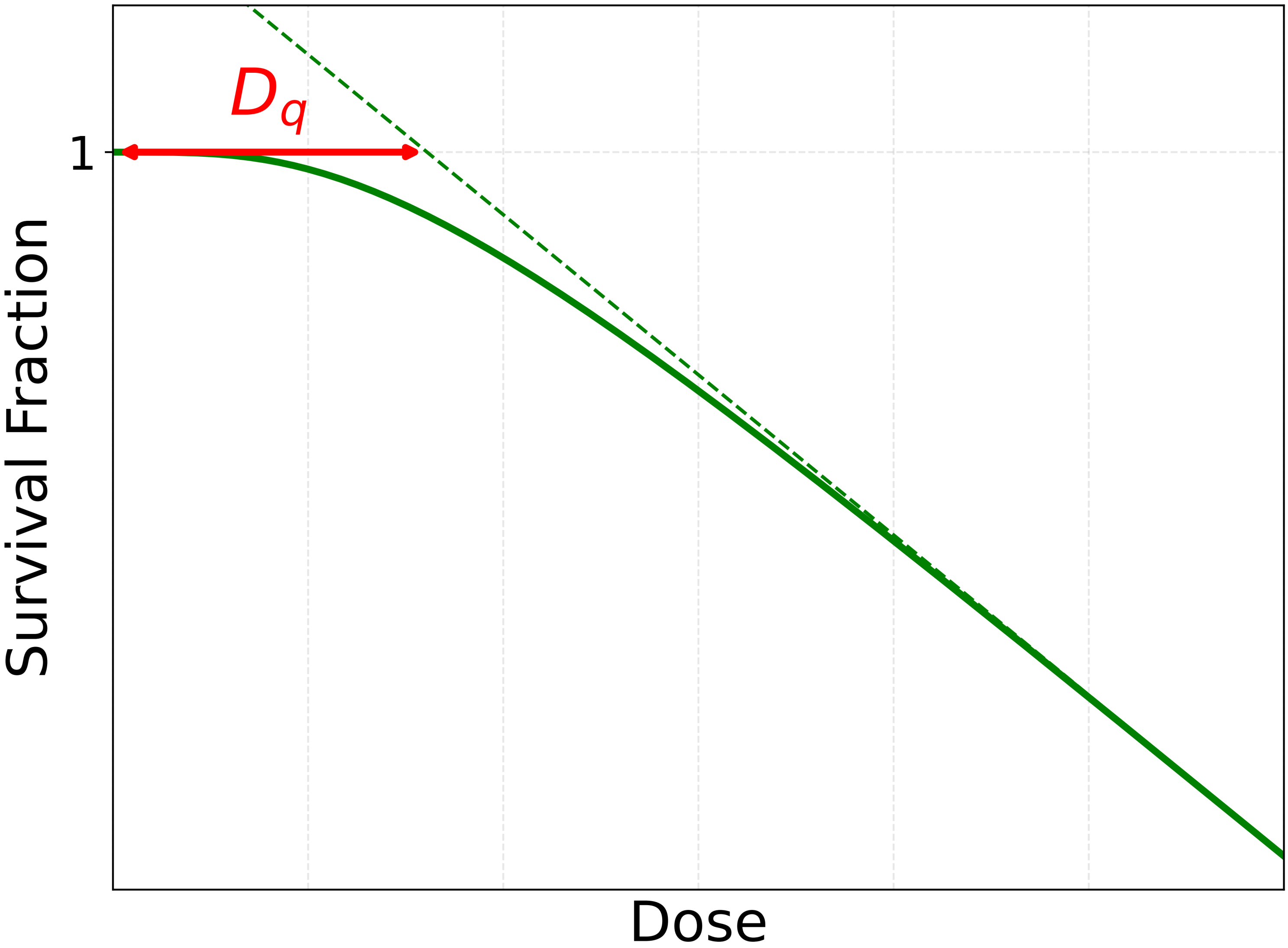

Figure 7

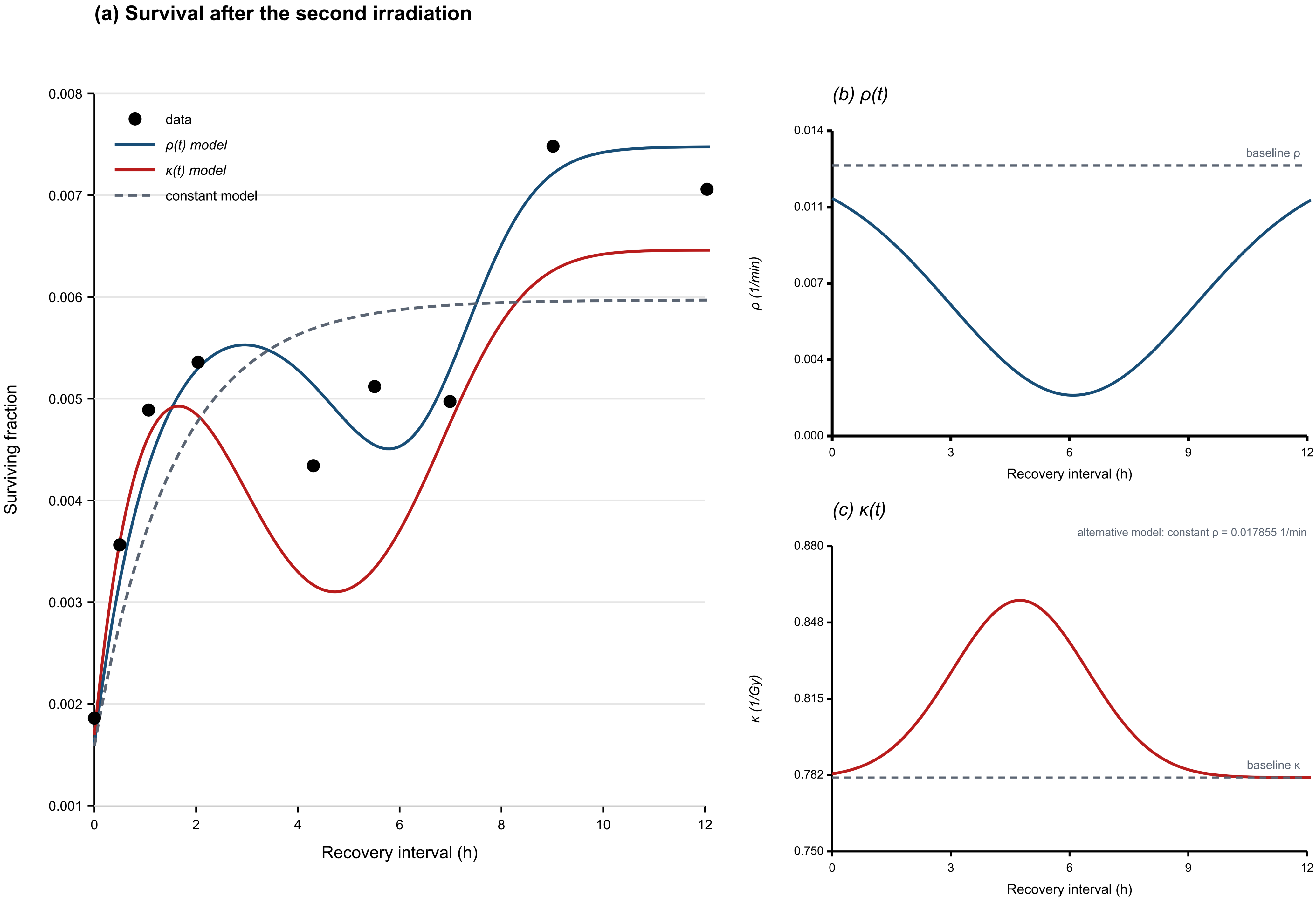

(a) Survival after the second irradiation
data
ρ(t) model
κ(t) model
constant model
Surviving fraction
Recovery interval (h)
0.008
0.007
0.006
0.005
0.004
0.003
0.002
0.001
0
2
4
6
8
10
12
(b) ρ(t)
ρ (1/min)
baseline ρ
0.014
0.011
0.007
0.004
0.000
0
3
6
9
12
Recovery interval (h)
(c) κ(t)
alternative model: constant ρ = 0.017855 1/min
κ (1/Gy)
baseline κ
0.880
0.848
0.815
0.782
0.750
0
3
6
9
12
Recovery interval (h)